\documentclass[twocolumn,reprint,amsmath,amssymb,cha,aip]{revtex4-1}

\usepackage{docs}
\usepackage{epsfig}% Include figure files
\usepackage{graphicx}% Include figure files
\usepackage{bm}%
\usepackage[colorlinks=true,linkcolor=blue]{hyperref}%

\begin{document}

%Title of paper
\title{Identification of the surface features in the electronic structure of Cr}

\author{Khadiza Ali, Shyama R. Varier, Deepnarayan Biswas,
Srinivas C. Kandukuri, and Kalobaran Maiti\footnote{Corresponding
author: kbmaiti@tifr.res.in}}

%\altaffiliation{Corresponding author: kbmaiti@tifr.res.in}

\address{Department of Condensed Matter Physics and Materials
Science, Tata Institute of Fundamental Research, Homi Bhabha Road,
Colaba, Mumbai - 400 005, INDIA.}

\date{\today}

\begin{abstract}
We studied the electronic structure of high quality Cr(110) films
grown on W(110) surface employing photoemission spectroscopy.
Experiments on the differently aged samples revealed distinct
signatures of the surface and bulk features in the electronic
structure. It is observed that the adsorbed oxygens form covalent
bonds with the surface Cr atoms at higher temperatures, while they
remain almost unreacted at low temperatures. In addition to the spin
density wave transition induced band folding across the bulk
Ne\'{e}l temperature, we discover a weakly dispersing sharp feature
emerging near the Fermi level at low temperatures presumably due to
correlation induced effects.
\end{abstract}

%\pacs{75.45.+j, 75.30.Fv, 75.30.Mb, 79.60.Bm}

\maketitle

%\section{Introduction}

Unusual magnetism of elemental Cr led to its wide ranging
applications involving recording media, high density storage media,
magnetic sensors, giant magnetoresistance based devices etc.
Enormous effort has been put forward to understand the exotic
electronic properties of this system, which is crucial to have
further developments in technological application of this material
as well as involved fundamental science. Bulk Cr forms in bcc
structure and is a good example of Fermi surface nesting driven
antiferromagnet \cite{Fawcett,review1,review2} exhibiting
incommensurate spin density wave (ISDW) transition at 311 K that
becomes commensurate via a spin-flip transition around 150 K.
Various studies revealed several controversies in its electronic
properties; for example, antiferromagnetic/ferromagnetic surface on
antiferromagnetic bulk \cite{review1,review2,gewinner,klebanoff},
complex surface-bulk differences in the electronic structure
\cite{klebanoff,ganesh,nakajima,donath}, orbital Kondo effect vs.
shockley surface states \cite{donath,Yu1,Yu2,wiesendanger},
proximity to quantum criticality \cite{yeh} etc. Contrasting
scenario on the energy gap has also been reported such as signature
of pseudogap \cite{pepin}, direct multiple gaps \cite{boekelheide1}
in the ISDW phase. Manifestation of such multifaceted exoticity in a
simple elemental system is remarkable and has continued to attract
much attention in the fundamental science and technology for long.

Device fabrication based on such materials requires knowledge of the
surface behavior - whether the bulk properties survives at the
surface, if differs how much is the deviation, etc. In order to
investigate the surface electronic structure critically, we studied
the evolution of the electronic structure of Cr with temperature and
aging. We prepared Cr(110) surface, which is antiferromagnetic and
expected not to favor orbital Kondo resonance. This allows us to
investigate the surface-bulk differences in the electronic structure
solely due to surface termination induced effect without major
change in their magnetism. Our results reveal surface and bulk
character of various photoemission spectral features unambiguously
and interesting evolution of the surface states on aging. Electron
correlation is found to play important role in the electronic
properties of this system.

%\section{Experimental details}

The sample was prepared in an ultrahigh vacuum chamber (base
pressure better than 1$\times$10$^{-10}$ Torr) by electron beam
evaporation of Cr onto a clean W(110) surface. The details of the
sample preparation and characterization is given in the
supplementary materials. Angle resolved photoemission spectroscopic
(ARPES) measurements were carried out using a Phoibos150 analyzer
from Specs GmbH and monochromatic He {\scriptsize I} source ($h\nu$
= 21.2 eV) with the momentum resolution fixed at
0.01$\times$2$\pi/a$, where the lattice constant of Cr, $a$ is
2.88~\AA. The $x$-ray photoemission spectroscopy (XPS) measurements
were carried out with a monochromatic Al $K\alpha$ source. All these
experiments were carried at a base pressure of about
1$\times$10$^{-10}$ Torr. The experiment temperature was achieved
using an open cycle helium cryostat, LT-3M from Advanced Research
Systems, USA.

%\section{Results}

\begin{figure}
 \vspace{-2ex}
\includegraphics [scale=0.4]{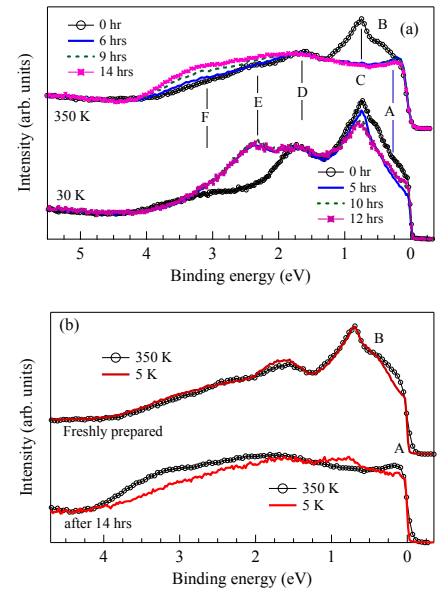}
 \vspace{-10ex}
\caption{(a)Aging of the normal emission spectra at 350 K and 30 K.
(b) Normal emission spectra at 350 K and 5 K from freshly prepared
sample (upper panel). The lower panel shows 14 hours aged spectrum
at 350 K (open circles) and then cooled the same aged sample to 5 K
(line).}
 \vspace{-2ex}
\end{figure}

First we investigate the evolution of the electronic structure with
aging. Angle integrated photoemission technique was employed with an
acceptance angle of $\pm$15$^o$ to enable quick collection of the
data with good signal to noise ratio. The spectra collected at 350 K
\& 30 K from freshly prepared sample are shown in Fig. 1(a) after
normalizing by the intensity around 6 eV binding energy. Each
spectrum exhibits several features at the binding energies 0.2, 0.4,
0.7, 1.5, 2.4 \& 3.2 eV - the features are denoted by A, B, C, D, E
\& F in the figure. The intensity between 0.3-1.2 eV binding energy
range in the 350 K spectra decreases dramatically with time and
becomes almost saturated within about 6 hours. Such spectral change
with aging can be attributed to the suppression of intensities by
the impurities adsorbed on the surface and/or bonded to the surface
atoms. The intensities near the Fermi level, $\epsilon_F$ appear to
remain uninfluenced by aging at 350 K. Curiously, a substantial
decrease in intensity of A is observed at 30 K. Since aging
influences the surface states most prominently, the feature A must
be possessing significant surface character at 30 K. Such phenomena
suggest a surface spectral weight transfer from higher binding
energy regime to the vicinity of $\epsilon_F$ with the decrease in
temperature and hence, the emergence of the surface character of the
intensities around 0.2 eV at low temperatures.

Subsequent to the decrease of B and C in the 350 K spectra, the
intensities of the features, E and F increase with aging. Employing
XPS (shown in the supplementary data) we observed that the intensity
of the oxygen 1$s$ signal grows with time delay as often observed by
us in various other studies \cite{bi2se3}. Therefore, the spectral
evolution observed here are attributed to the influence of oxygens
on the surface states. The feature, E appears due to the non-bonding
oxygen levels, and the features D \& F correspond to the energy
bands hybridized to oxygen 2$p$ states. The scenario at 30 K is
significantly different; in addition to large decrease in intensity
in the vicinity of $\epsilon_F$, intensity of the feature E becomes
significantly intense within a very short time delay keeping the
features D and F intensities almost unchanged. This suggests that
the Cr-O bonding is less significant at low temperatures and the
presence of large amount of adsorbed oxygen that leads to an
enhancement of the non-bonding feature, E.

The temperature evolution of the electronic structure across the
bulk Ne\'{e}l temperature of 311 K is investigated in Fig. 1(b). The
350 K and 5 K spectra obtained from the freshly prepared sample are
compared in the upper panel of Fig. 1(b). Distinct decrease in
intensity in the vicinity of $\epsilon_F$ is observed in addition to
a small increase in intensity of the feature D. The surface features
remain unchanged in this large temperature range due to the fact
that the surface magnetic transition occurs at a higher temperature
and the changes within the magnetically ordered phase is negligible
\cite{klebanoff,Gd}. This is further verified by cooling down the 14
hours aged sample possessing primarily the bulk features near
$\epsilon_F$. The 350 K and 5 K spectra exhibit significant decrease
in intensity near $\epsilon_F$ as a signature of antiferromagnetic
transition \cite{raviAPL}. Interestingly, the intensity around 0.7
eV (feature C) becomes stronger along with a decrease in intensity
beyond 1.5 eV binding energy. This corroborates the conclusions of
less Cr-O bonded objects at 5 K due to less reactivity of oxygens at
lower temperatures.

\begin{figure}
 \vspace{-2ex}
\includegraphics [scale=0.4]{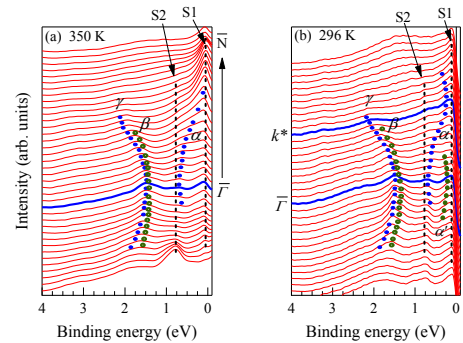}
 \vspace{-36ex}
\caption{ARPES data at (a) 350 K and (b) 296 K. symbols show the
peak positions.}
 \vspace{-2ex}
\end{figure}

In order to investigate the band character of the above features,
the energy band dispersions of Cr(110)/W(110) sample obtained by
ARPES are shown in Figs. 2(a) and 2(b). Distinct signature of three
energy bands denoted by $\alpha$, $\beta$ and $\gamma$ are observed.
These three energy bands possess behavior quite similar to the
$t_{2g}$ bands obtained from band structure calculations
\cite{bandstr1a,bandstr1b,bandstr2}; the degeneracy of the $t_{2g}$
bands at $\Gamma$ is lifted due to the fact that He {\scriptsize I}
photon energy ($h\nu$ = 21.2 eV) corresponds to a $k_z$-value
($k_z\sim$1.7$\times$2$\pi\over c$; $c$ is the lattice constant
along (110) direction) away from the high symmetry point, $\Gamma$.
Such photon energy dependence of the band dispersion will be absent
for the surface bands, which is the primary focus of this study.
There could be additional effect due to the strain arising from the
differences in the lattice constants of Cr films with the W(110)
substrate. The energy dispersion of the $\alpha$ band with $d_{xy}$
($z$-axis along the surface normal) symmetry makes an electron
pocket around the $\Gamma$-point and resembles well to the
theoretically calculated results
\cite{bandstr1a,bandstr1b,bandstr2}.

In addition to the bulk bands, distinct signature of a weakly
dispersive band is observed around 0.7 eV binding energy denoted by
S2 in the figure, which is within the energy gap of the bulk bands
$\alpha$ and $\beta$. Another almost non-dispersive feature denoted
by S1 also appears near $\epsilon_F$. These features are found to be
significantly sensitive to aging as shown in Fig. 1 (features, A and
C) indicating their surface character. The comparison of the
experimental spectra at 296 K and 350 K in Figs. 2(a) and 2(b)
reveals interesting evolution. While all the energy bands remain
almost similar, the $\alpha$ band appears to fold back at $k^*$ as
shown in the figure - signature of band folding due to
incommensurate SDW transition. In addition, a new feature denoted by
$\alpha^\prime$ appears near $\bar{\Gamma}$.

\begin{figure}
 \vspace{-2ex}
\includegraphics [scale=0.4]{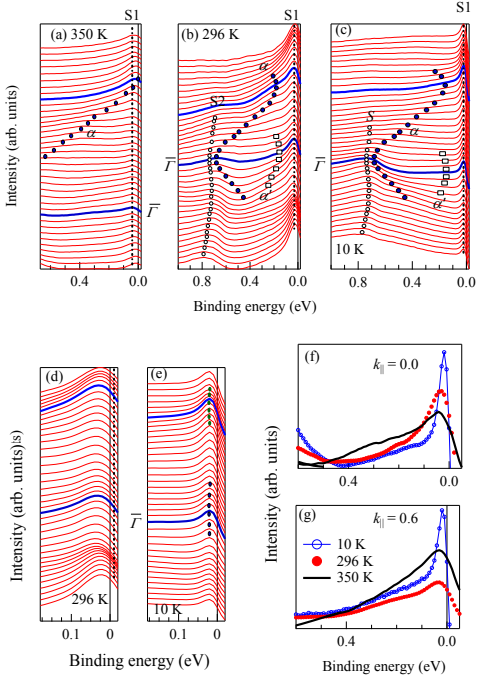}
 \vspace{-6ex}
\caption{Experimental energy bands at (a) 350 K, (b) 296 K and (c)
10 K. The energy distribution curves (EDC) with expanded energy
scale at 296 K and 10 K are shown in (d) and (e), respectively. EDCs
at different temperatures at (f) $k_{||} = 0.0$ and (g) $k_{||} =
0.6$.}
 \vspace{-2ex}
\end{figure}

The temperature evolution of the features near $\epsilon_F$ as a
function of temperature is shown in Fig. 3. In addition to the band
folding effect, the comparison of the spectra at 350 K, 296 K and 10
K exhibit emergence of a sharp feature around 15 meV below
$\epsilon_F$ possessing weak dispersion. The distinct nature of this
additional intensity is most evident in the energy distribution
curves shown in Fig. 3(f) and 3(g). Moreover, it exhibits different
trend in the change in intensities with temperature at $\Gamma$ and
$k^*$ ($k_{||} = 0.6$) - the intensity at $\epsilon_F$ gradually
increases with the decrease in temperature at $\Gamma$ while the
intensity at 296 K is the lowest at $k^*$.

\begin{figure}
 \vspace{-2ex}
\includegraphics [scale=0.4]{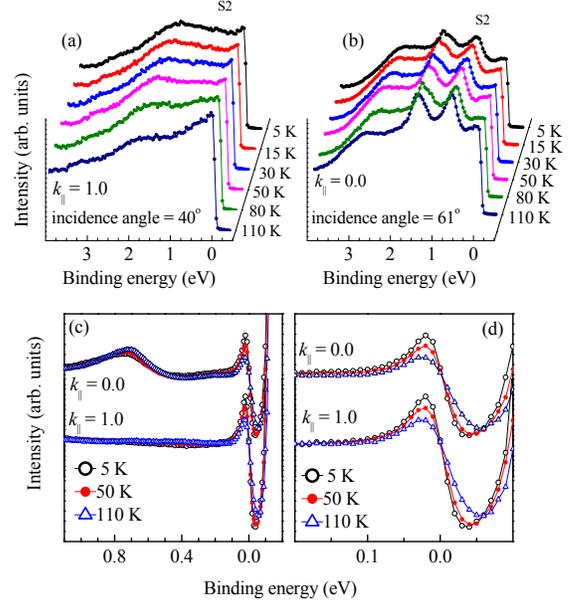}
 \vspace{-6ex}
\caption{Temperature evolution of EDC at (a) $k_{||} = 1.0$ and (b)
$k_{||} = 0.0$. (c) and (d) show the spectral density of states near
$\epsilon_F$ at different temperatures in an expanded energy scale.}
 \vspace{-2ex}
\end{figure}

Detailed temperature evolution of the features in the low
temperature range (below 110 K) are shown in Fig. 4, where we show
the energy distribution curves at $\bar{\Gamma}$ and $k_{||}$ = 1.0.
The features around 1.5 eV binding energy represent the intensities
of the $\beta$ \& $\gamma$ bands corresponding to the bulk
electronic structure and exhibit enhancement in intensity with the
decrease in temperature at $k_{||}$ = 1.0, while it seems opposite
at $\bar{\Gamma}$. Overall, the intensity increases at 1.5 eV as
observed in the angle integrated spectra shown in Fig. 1(b)
suggesting an enhancement of the local moment of the system within
the magnetically ordered phase. The intensities at $\epsilon_F$
exhibit significant enhancement at low temperatures and asymmetry
with respect to $\epsilon_F$ \cite{raviEPL}. This is shown in Figs.
4(c) and 4(d) after dividing the experimental spectra by the
resolution broadened Fermi-distribution function.

%\section{Discussion}

It is clear that the electronic structure of one of the most used
elemental metal, Cr is quite complex. The results of our extensive
experimental study unveil plethora of interesting scenarios, which
have significant impact on application of this materials for future
technological advancements. Firstly, we identify the surface
features via reduction in intensity due to aging and observe that
the degradation of the surface features of Cr film strongly depends
on temperature. At low temperatures, we observe signature of
adsorbed oxygen, which reduces the surface feature intensities. At
elevated temperatures, the oxygens form bonding with the surface Cr
atoms that enhances Cr spectral contributions at higher binding
energies (around 3.2 eV) due to the bonding-antibonding splitting of
the electronic states. The enhancement of the reactivity to surface
impurities at higher temperature happens due to the thermal energy
providing suitable activation for such reactions.

While the above results on aging of the Cr surface provide details
of the surface stability, the spectral evolution due to
adsorption/bonding of impurities with the surface atoms helps to
identify the surface and bulk features in the electronic structure.
There have been controversy on the surface character of the spectral
feature at 0.7 eV binding energy denoted by S2 in Fig. 1.
Substantial decrease in intensity with aging unambiguously indicates
surface character of S2. The other important issue involves the
observation of significant sensitivity of S1 to aging at 30 K and
it's absence at 350 K. This suggests that the predominant surface
character around 0.2 eV becomes evident at low temperatures
presumably due to the fact that the bulk contributions near
$\epsilon_F$ diminishes significantly at low temperatures due to the
formation of SDW gap making the relative surface contributions near
$\epsilon_F$ discernible at low temperatures. In addition, there
could be surface spectral weight transfer with the decrease in
temperature.

Photoemission measurements at room temperature with varied angle of
incident/emitted photon beam concluded Shockley type surface states
with $d_{z^2}$ character of S1\cite{donath}. In order to learn the
orbital character of the growing intensities at low temperatures, we
compare the spectra obtained at two geometries - the incident photon
beam at 61$^o$ ($k_{||}$=0.0) and 40$^o$ ($k_{||}$=1.0) with respect
to the surface normal. Thus, the $k_{||}$ = 1.0 case becomes
slightly more favorable to probe $d_{xz}d_{yz}$ states and the
$d_{z^2}$ states becomes significantly less sensitive due to
relatively more in-plane alignment of the light polarization vector.
In Fig. 4(d), we observe that the intensity of the sharp feature at
15 meV becomes more intense in the $k_{||}$=1.0 spectra suggesting
$d_{xz}d_{yz}$ character of the growing feature over the intensities
of S1.

It is to note here that the studied Cr(110) surface is unlikely to
exhibit orbital Kondo behavior as the surface is expected to be
antiferromagnetic, while orbital Kondo effect corresponds to
spin-ferromagnetic order \cite{orbKondo}. In general, correlated
electron systems such as rare-earths \cite{rb6} and transition metal
oxides \cite{RMP} exhibit temperature reduction induced growth of
the coherent feature at $\epsilon_F$ with respect to the intensities
of the incoherent feature representing the correlation induced
electronic states. The emergence of the sharp feature around 15 meV
binding energy with the decrease in temperature observed here
presumably suggests similar scenario in this system. Clearly, more
studies are necessary to understand the complexity of the electronic
structure of this system.

%\section{Conclusions}

In summary, we have investigated the detailed electronic structure
of high quality Cr(110) films grown on W(110) employing angle
resolved photoemission spectroscopy. The aging of the sample surface
helped to reveal the surface and bulk character of various spectral
features. The adsorbed oxygens on the surface form bonds with the
surface Cr atoms at temperatures close to room temperature, while
they behave like adsorbed gas at low temperatures. The bulk
electronic structure exhibits signature of band folding due to bulk
spin density wave transition. There are two surface peaks around 0.2
eV and 0.7 eV binding energies. The temperature variation down to 5
K reveals emergence of an additional sharp feature corresponding to
the surface electronic structure. These results reveal the complex
surface behavior of Cr, instability of the surface states with aging
and the importance of electron correlation induced effect in the
electronic structure, which are important to understand for its
potential applications.

\section{Acknowledgements}

The authors, S. R. V. and K. M. acknowledge financial support from
the Dept. of Science and Technology, Govt. of India under the
Swarnajayanti fellowship programme.


\begin{thebibliography}{99}
%
\bibitem{Fawcett}
E. Fawcett, Rev. Mod. Phys. {\bf 60,} 209 (1988).
%
\bibitem{review1}
F. Schiller, D. V. Vyalikh, V. D. P. Servedio and S. L. Molodtsov,
Phys. Rev. B {\bf 70,} 174444 (2004);
%
\bibitem{review2}
E. Rotenberg {\it et al.}, New J. Phys. {\bf 7,} 114 (2005); E.
Rotenberg {\it et al.}, New J. Phys. {\bf 10,} 023003 (2008).
%
\bibitem{gewinner}
G. Gewinner, J. C. Peruchetti, A. Ja\'{e}gl\'{e} and R. Pinchaux,
Phys. Rev. B {\bf 27,} 3358 (1983).
%
\bibitem{klebanoff}
L. E. Klebanoff, S. W. Robey, G. Liu and D. A. Shirley, Phys. Rev. B
{\bf 30,} 1048(R) (1984).
%
\bibitem{ganesh}
G. Adhikary, R. Bindu, S. Patil and K. Maiti, Appl. Phys. Letts.
{\bf 100}, 042401 (2012); G. Adhikary, R. Bindu, S. K. Pandey and K.
Maiti, Europhys. Letts. {\bf 99}, 37009 (2012).
%
\bibitem{nakajima}
N. Nakajima, O. Morimoto, H. Kato and Y. Sakisaka, Phys. Rev. B {\bf
67}, 041402(R) (2003).
%
\bibitem{donath}
M. Budke, T. Allmers, M. Donath and M. Bode, Phys. Rev. B {\bf 77,}
233409 (2008).
%
\bibitem{Yu1}
O. Yu. Kolesnychenko {\it et al.}, Nature {\bf 415}, 507 (2002).
%
\bibitem{Yu2}
O. Yu. Kolesnychenko, Phys. Rev. B {\bf 72}, 085456 (2005).
%
\bibitem{wiesendanger}
T. H\"{a}nke {\it et al.}, Phys. Rev. B {\bf 72}, 085453 (2005).
%
\bibitem{yeh}
A. Yeh {\it et al.}, Nature {\bf 419}, 459 (2002).
%
\bibitem{pepin}
C. P\'{e}pin and M. R. Norman, Phys. Rev. B {\bf 69}, 060402(R)
(2004);
%
%\bibitem{Ganesh-AIP}
G. Adhikary, R. Bindu, S. Patil and K. Maiti, AIP Conf. Proc. {\bf
1349}, 819 (2011).
%
\bibitem{boekelheide1}
Z. Boekelheide, E. Helgren and F. Hellman, Phys. Rev. B {\bf 76},
224429 (2007).
%
%\bibitem{ARPES}
%J. Sch\"{a}fer, E. Rotenberg, S. D. Kevan and P. Blaha, Surf.
%Science {\bf 454}, 885 (2000).
%
%\bibitem{surface}
%K. Maiti {\it et al.}, Phys. Rev. B {\bf 73}, 052508 (2006); R. S.
%Singh, V. R. R. Medicherla, K. Maiti and E. V. Sampathkumaran, Phys.
%Rev. B {\bf 77}, 201102 (2008); K. Maiti, R. S. Singh and V. R. R.
%Medicherla, Phys. Rev. B {\bf 76}, 165128 (2007).
%
\bibitem{bi2se3}
D. Biswas, S. Thakur, K. Ali, G. Balakrishnan and K. Maiti,
Scientific Rep. {\bf 5} 10260 (2015); D. Biswas and K. Maiti,
Europhys. Lett. {\bf 110}, 17001 (2015).
%
\bibitem{Gd}
K. Maiti {\it et al.}, Phys. Rev. Lett. {\bf 88}, 167205 (1988); K.
Maiti {\it et al.}, Phys. Rev. Lett. {\bf 86}, 2846 (2001).
%
\bibitem{raviAPL}
R. S. Singh, V. R. R. Medicherla and K. Maiti, Appl. Phys. Lett.
{\bf 91}, 132503 (2007).
%
%\bibitem{wien2k}
%P. Blaha, K. Schwarz, G. K.H. Madsen, D. Kvasnicka and J. Luitz,
%{\bf WIEN2k}, {\it An Augmented Plane Wave + Local Orbitals Program
%for Calculating Crystal Properties} (Karlheinz Schwarz, Techn.
%Universit\"{a}t Wien, Austria), 2001. ISBN 3-9501031-1-2.
%
\bibitem{bandstr1a}
S. Asano and J. Yamashita, J. Phys. Soc. Jpn. {\bf 23}, 714 (1967).
%
\bibitem{bandstr1b}
R. H. Victora and L. M. Falicov, Phys. Rev. B {\bf 31}, 7335 (1985).
%
\bibitem{bandstr2}
P. Habibi, C. Barreteau and A. Smogunov, J. Phys. Condens. Matter
{\bf 25}, 146002 (2013).
%
%\bibitem{nesting1}
%E. Fawcett, Rev. Mod. Phys. {\bf 60}, 209 (1988).
%
%\bibitem{nesting}
%J. Graebner and J. A. Marcus, J. Appl. Phys. {\bf 37}, 1262 (1966).
%
\bibitem{raviEPL} K. Maiti, R. S. Singh, V. R. R. Medicherla,
EPL {\bf 78}, 17002 (2007).
%
\bibitem{orbKondo} D.L. Cox and A. Zawadowski, Adv. Phys. {\bf 47},
599 (1998).
%
\bibitem{rb6}
S. Patil, G. Adhikary, G. Balakrishnan and K. Maiti, J. Phys.
Condens. Matter {\bf 23}, 495601 (2011); S. Patil {\it et al.}, J.
Phys. Condens. Matter {\bf 22}, 255602 (2010); S. Patil {\it et
al.}, Phys. Rev. B {\bf 82}, 104428 (2010).
%
\bibitem{RMP} A. Georges, G. Kotliar, W. Krauth, and M.J. Rozenberg, Rev. Mod.
Phys. {\bf 68}, 13 (1996); M. Imada, A. fujimori, and Y. Tokura,
Rev. Mod. Phys. {\bf 70}, 1039 (1998).

%
\end{thebibliography}
\end{document}